\documentclass[a4paper]{article}

\usepackage{xcolor}
\usepackage{multirow}
\usepackage{INTERSPEECH2020}
\usepackage{amsmath,graphicx,multirow,array,tikz,subfig,bm,amsfonts}
\newcommand{\PreserveBackslash}[1]{\let\temp=\\#1\let\\=\temp}
\newcolumntype{C}[1]{>{\PreserveBackslash\centering}p{#1}}
\newcolumntype{R}[1]{>{\PreserveBackslash\raggedleft}p{#1}}
\newcolumntype{L}[1]{>{\PreservemBackslash\raggedright}p{#1}}

\title{ASAPP-ASR: Multistream CNN and Self-Attentive SRU\\for SOTA Speech Recognition}
\name{Jing Pan*, Joshua Shapiro*, Jeremy Wohlwend*, Kyu J. Han*, Tao Lei*\thanks{*Equal contributors.} and Tao Ma}
\address{
  ASAPP Inc.}
\email{\{jpan,jshapiro,jeremy,khan,tao,tma\}@asapp.com}

\begin{document}

\maketitle
\begin{abstract}
In this paper we present state-of-the-art (SOTA) performance on the LibriSpeech corpus with two novel neural network architectures, a \textit{multistream CNN} for acoustic modeling and a \textit{self-attentive simple recurrent unit (SRU)} for language modeling. In the hybrid ASR framework, the multistream CNN acoustic model processes an input of speech frames in multiple parallel pipelines where each stream has a unique dilation rate for diversity. Trained with the SpecAugment data augmentation method, it achieves relative word error rate (WER) improvements of 4\% on test-clean and 14\% on test-other. We further improve the performance via $N$-best rescoring using a 24-layer self-attentive SRU language model, achieving WERs of 1.75\% on test-clean and 4.46\% on test-other.   
\end{abstract}

\section{Introduction}
Deep neural networks (DNNs) have brought revolutionary changes to the landscape of speech recognition research. Since their advent \cite{Hinton12}, DNNs ranging from LSTMs \cite{Hochreiter97} to Transformers based on multi-headed self-attention \cite{vaswani} have contributed to inching the performance of speech recognition systems closer to human-level accuracy. 

In \cite{Xiong16-2} it was reported that human transcribers were given the evaluation data (also known as HUB5 eval2000) of the NIST 2000 Evaluation Challenge of Conversational Telephone Speech (CTS) \cite{William00}. Their average error rate on the Switchboard (SWBD) portion of the HUB5 eval2000 set was 5.9\%. In the same paper an ASR system was proposed, fusing variants of deep CNNs such as VGG \cite{simoyan14} or ResNet \cite{He16} with layer-wise context expansion and attention (LACE) \cite{yu16} as well as bi-directional LSTM (bLSTM) trained with the lattice-free MMI (LF-MMI) loss \cite{Povey16}. The system was claimed to have surpassed the human level of accuracy (5.8\% WER) on the SWBD eval set. In \cite{Saon17}, a similar experiment discovering the human ability for speech recognition was conducted, suggesting a new human-level error rate of 5.3\% on the SWBD eval set. Their proposed ASR system reached 5.5\% WER, combining ResNets and bLSTMs with speaker-adversarial multi-task learning. The WER on SWBD has been further pulled down to the state-of-the-art (SOTA) level of 5.1\% thanks to various novel DNN architectures, such as CNN-bLSTMs \cite{Han17,Xiong17}, highway LSTMs \cite{Kurata17}, and densely connected LSTMs \cite{han18,capio}. 

Another active area of research in speech recognition focuses on the well-known LibriSpeech corpus \cite{panayotov2015librispeech} where roughly 1,000hrs of spoken utterances were collected from audio books. Unlike the SWBD evaluation challenges, a number of end-to-end (E2E) ASR systems have been competitive on the LibriSpeech test sets, even exceeding the performance of hybrid HMM/DNN ASR systems. Listen, Attend and Spell (LAS) \cite{chan}, which uses a sequence-to-sequence architecture is the most representative E2E model for ASR. It consists of a pyramidal structure of bLSTMs for the encoder with content-aware attention. Using the SpecAugment method for data augmentation \cite{specaugment}, the LAS-based ASR system surpassed the accuracy on the test sets of LibriSpeech, boasting the WERs of 2.5\% and 5.8\% on test-clean and test-other, respectively.

More recent systems, both E2E and hybrid, utilized improvements from powerful Transformer models \cite{vaswani} for both AM and LM \cite{Luscher2019,han19-2,synnaeve19,wang20}. In \cite{Luscher2019}, a hybrid AM with sequence discriminative training was boosted by Transformer LM rescoring. \cite{han19-2} applied a multistream architecture to the hybrid ASR setting, where in each stream the multi-headed self-attention layer, following the shallow layers of TDNN-F \cite{povey18tdnnf}, was modified with factorizing the feed-forward layer inside. In \cite{synnaeve19}, E2E Transformer models with a sequence-to-sequence loss were trained for both AM and LM, and the outputs of the E2E models were rescored with a Transfomer LM as well as a gated CNN (GCNN) LM \cite{dauphin17}. Interestingly, 60k hours of extra audio data \cite{librilight,librivox14} were used for the semi-supervised AM update to further boost the accuracy of the proposed system. In \cite{wang20}, a deep Transformer architecture was analysed for AM with an iterated loss \cite{Tjandra20} in the hybrid ASR framework. Recently, a CNN-RNN-Transducer architecture was introduced \cite{han2020contextnet}, demonstrating even further improvement using Transformer models on the LibriSpeech benchmark. 

This paper presents new benchmark results for test-clean and test-other in LibriSpeech, 1.75\% and 4.46\%, respectively, thanks to the novel neural network architectures of AM and LM in \textit{multistream CNN} \cite{han20} and \textit{self-attentive simple recurrent unit (SRU)}. 
The multistream CNN acoustic model, inspired by \cite{han19-2} but without the multi-headed self-attention layers, processes input speech frames in multiple parallel pipelines where each stream has a unique dilation rate for the convolution kernels of CNNs for diversity. Trained with SpecAugment, it achieves relative WER improvements of 4\% on test-clean and 14\% on test-other. We further improve the performance with $N$-best rescoring using a 24-layer self-attentive SRU language model. SRU was proposed in \cite{lei18} for higher parallelization in recurrence computation. Our variant adds self-attention to the original SRU to not only replace some of linear operations in computation but also enhance context modeling capability. We rescore the $N$-best outputs of the lattices once rescored with the TDNN-LSTM language model trained by the Kaldi toolkit \cite{Povey11,Li2018}. The average relative WER improvement by the self-attentive SRU LM is around 23\% on both of the test sets. 

This paper is organized as follows. In Section 2, we describe the details of our ASAPP-ASR system focusing on the neural network architectures for AM and LM as well as the LM rescoring strategies leveraged. In Section 3, we provide the experimental setup and discuss the results from various configurations in the proposed system. In Section 4, we conclude the paper with summary remarks and future directions.

\begin{figure}[t]
  \centering
  \includegraphics[height=8.5em, width=\linewidth]{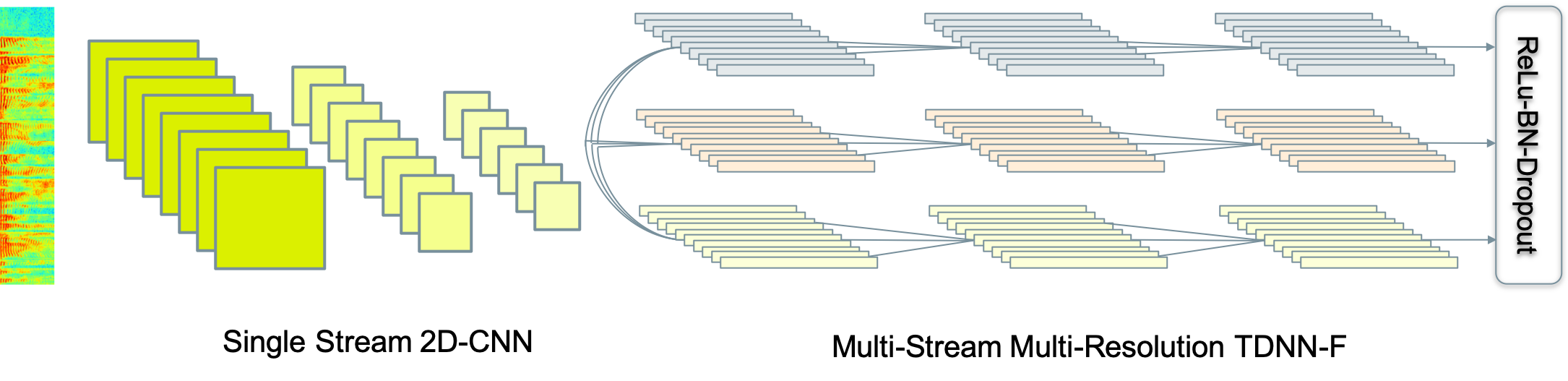}
  \caption{Schematic diagram of the multistream CNN acoustic model architecture.}
  \label{fig:speech_production}
\end{figure}

\section{ASAPP-ASR: System Descriptions}

\subsection{Multistream CNN for Acoustic Modeling}
For robust acoustic modeling, we leverage the benefits of multistream CNNs \cite{han20} (illustrated in Figure 1 above). This novel neural network architecture accommodates diverse temporal resolutions in multiple streams to achieve robustness. For diverse temporal resolution, it considers stream-specific dilation rates on TDNN-F \cite{povey18tdnnf}, a variant of 1D-CNN. Each stream stacks narrower TDNN-F layers whose kernel has a unique dilation rate when processing input speech frames in parallel. The dilation rate for the TDNN-F layers in each stream is chosen from multiples of the default subsampling rate (3 frames). This offers a seamless integration with the training and decoding process where input speech frames are subsampled. With SpecAugment \cite{specaugment}, multistream CNNs provide additional robustness against challenging audio, such as the ``other'' sets in LibriSpeech. 

In the proposed architecture, we position 5 layers of 2D-CNNs in a single stream fashion to process input log-mel spectrogram before multiple streams are branched out. We use $3 \times 3$ kernels for the 2D CNN layers with a filter size of 256 except for the first layer with a filter size of 128. Every other layer in the 2D-CNNs we apply frequency band subsampling with a rate of 2. In the multistream structure, each stream rolls out 17 TDNN-F layers, where each TDNN-F consists of two $2 \times 1$ factorized convolution matrices, followed by a skip connection, batch normalization and dropout layer, with 512-dimensional neurons. Consider the embedding vector $\mathbf{x}_i$ coming out of the single stream 2D-CNN layers at the given time step of $i$. The output vector $\mathbf{y}^m_i$ from the stream $m$ going through the stack of TDNN-F layers with the dilation rate $r_m$ can be written as
\begin{align}
    \mathbf{y}^m_i = \textit{Stacked-TDNN-F}_m \left( \mathbf{x}_i | \left[ -r_m, r_m \right] \right)
\end{align}
where $[ -r_m, r_m]$ means a $3 \times 1$ kernel given the dilation rate of $r_m$. The output embeddings from the multiple streams are then concatenated, and followed by ReLu, batch normalization and a dropout layer:
\begin{align}
    \mathbf{z}_i = \textit{Dropout} \left( \textit{BN} \left( \textit{ReLu} \left( \textit{Concat} \left( \mathbf{y}^1_i, \mathbf{y}^2_i, \dots, \mathbf{y}^M_i \right) \right) \right) \right).
\end{align}
This embedding vector is projected on the output layer via a couple of fully connected layers at the end of the network. We employ 3 streams with the dilation configuration of \texttt{6-9-12} where the dilation rates of 6, 9 and 12 are applied for TDNN-F layers across the 3 streams.

\subsection{Self-attentive SRU for Language Modeling}
We train our LMs using a variant of the SRU architecture proposed by \cite{lei18}. 
Given an input sequence $\{\mathbf{x}_1, \cdots, \mathbf{x}_L\}$ where each $\mathbf{x}_t \in \mathbb{R}^d$ represents a feature vector, a single layer of SRU involves the following recurrence computation:
\begin{align*}
    \mathbf{f}_t &= \sigma(\mathbf{u}_{1,t} + \mathbf{v}\odot \mathbf{c}_{t-1} + \mathbf{b}) \\
    \mathbf{r}_t &= \sigma(\mathbf{u}_{2,t} + \mathbf{v}'\odot \mathbf{c}_{t-1} + \mathbf{b}') \\
    \mathbf{c}_t &= \mathbf{f}_t\odot \mathbf{c}_{t-1} + (1-\mathbf{f}_t)\odot \mathbf{u}_{3,t} \\
    \mathbf{h}_t &= \mathbf{r}_t\odot \mathbf{c}_t + (1-\mathbf{r}_t) \odot \mathbf{x}_t .
\end{align*}
where $\sigma( \cdot )$ is the sigmoid activation, $\mathbf{h}_t \in \mathbb{R}^d$ is the output state at step $t$, and $\mathbf{c}_t, \mathbf{f}_t, \mathbf{r}_t\in \mathbb{R}^d$ are the internal hidden state and sigmoid gates at step $t$, respectively.
The vectors $\mathbf{u}_{*,t}$ are computed using a linear projection:
\begin{align}
\mathbf{u}_{1,t},\, \mathbf{u}_{2,t},\, \mathbf{u}_{3,t} = \left[\mathbf{W}_1,\, \mathbf{W}_2,\, \mathbf{W}_3\right]^\top \mathbf{x}_t,
\label{eq:u_proj}
\end{align}
given three parameter matrices $\mathbf{W}_1$, $\mathbf{W}_2$, $\mathbf{W}_3$ of the SRU layer. 
Compared to other recurrent networks such as LSTM and GRU, SRU adopts element-wise hidden-to-hidden connections $\mathbf{v}\odot \mathbf{c}_{t-1}$ and $\mathbf{v}'\odot \mathbf{c}_{t-1}$.
As a consequence, each of the hidden dimensions becomes independent and can be executed in parallel, achieving much faster training speed.

Several variants of SRU architecture have been successfully employed in speech models~\cite{park2018fully,shangguan2019optimizing,Koriyama_2020}.
In this work, we use a self-attentive variant to enhance context modeling capacity by substituting the linear operation (Eq. \ref{eq:u_proj}) with the multi-head attention operation originally proposed in \cite{vaswani}.

\subsection{Language Model Rescoring}
We employ multiple stages of LM rescoring in order to obtain the minimum WERs. The initial decoding is based on the decoding graph constructed from the multistream CNN AM and a 3-gram LM, resulting in the initial hypotheses in a lattice format. Lattice rescoring is done with a larger sized 4-gram LM, followed by a second-pass lattice rescoring with the TDNN-LSTM language model \cite{Li2018}. In the final rescoring stage, we use an interpolated self-attentive SRU LM. We linearly interpolate two self-attentive SRU models, one of which is trained on word pieces using byte-pair encoding (BPE) and the other is trained at the word level. With this interpolation, we re-rank the $N$-best hypotheses from the lattices rescored by the TDNN-LSTM LM in the previous stage. In our experiments, we empirically keep $N$ at 100. 

In the final stage of rescoring, the $N$-best hypotheses are re-ranked by the combination of an estimated AM and LM likelihood for a hypothesized word sequence $\mathcal{S}$ given acoustic features $\mathcal{O}$,
\begin{align}
    P(\mathcal{S}|\mathcal{O}) \approx P_{AC}(\mathcal{S}|\mathcal{O})P_{LM}(\mathcal{S})^{\lambda_{\alpha}}
\end{align}
where $P_{AC}(\mathcal{S}|\mathcal{O})$ is the AM likelihood estimate, $P_{LM}(\mathcal{S})$ is the LM likelihood estimate, which can be further detailed as below,
\begin{align}
    P_{LM}(\mathcal{S}) = P_{SRU^*}(\mathcal{S})^{\lambda_{\beta}}P_{TL}(\mathcal{S})^{1-\lambda_{\beta}}
\end{align}
where $P_{SRU^*}(\mathcal{S})$ is the likelihood estimate from the interpolated SRU LM given an interpolation weight $\lambda_{\gamma}$ for the BPE SRU model (i.e., $1-\lambda_{\gamma}$ for the word SRU LM), $P_{TL}(\mathcal{S})$ is the likelihood estimate from the TDNN-LSTM LM, and $\lambda_{\alpha}$, $\lambda_{\beta}$ and $\lambda_{\gamma}$ are the hyper parameters which we optimize through a grid search on the dev data in LibriSpeech.

We further refine the ranking with the minimum expected word error objective \cite{stolcke1997explicit}, defined by:
\begin{align}
    E[err(\mathcal{S})|\mathcal{O}] \approx \sum_{i=1}^{N}\bar{P}(\mathcal{S}_i|\mathcal{O})err(\mathcal{S}|\mathcal{S}_i)
\end{align}
where $\bar{P}(\mathcal{S}_i|\mathcal{O}) = P(\mathcal{S}_i|\mathcal{O}) / \sum_{j=1}^{N}P(\mathcal{S}_j|\mathcal{O})$ is the normalized term of $P(\mathcal{S}_i|\mathcal{O})$ and $err(\mathcal{S}|\mathcal{S}_i)$ is the WER measure of $\mathcal{S}$ with $\mathcal{S}_i$ as reference. Applying the expected word error minimization can weaken a bias on the sentence-level likelihood maximization and shift the ranking focus toward the local word-level accuracy. In order to reduce the computation complexity, we first rank the $N$-best hypotheses by the sentence-level likelihood maximization (Eq. 4), and then update the rank of the top 20 hypotheses by the minimum expected word error (Eq. 6).

\section{Experimental Setup and Results}

\subsection{LibriSpeech}
We conduct the experiments on the LibriSpeech corpus \cite{panayotov2015librispeech}, which is a collection of approximately 1,000hr read speech (16kHz) from the audio books that are part of the LibriVox project \cite{librivox14}. The training data is split into 3 partitions of 100hrs, 360hrs, and 500hrs while both of the dev and test data are split into `clean’ and `other’ categories, where each category contains around 5hrs of audio. The corpus provides extra written texts of 800M words\footnote{http://openslr.org/11.} for LMs. We normalize them to correct typos as well as spelling consistencies between British and American English. The same normalization is applied to all the text transcripts of the training, dev and test set to be consistent.

\subsection{Acoustic Models and Non-SRU Language Models}
We follow the conventional steps to train hybrid GMM/HMM acoustic models using the default Kaldi recipe for LibriSpeech\footnote{https://github.com/kaldi-asr/kaldi/tree/master/egs/librispeech/s5.}, up to a point where a triphone model is trained with speaker-adaptive training (SAT) with feature-space MLLR (fMLLR) to further refine Gaussian mixture parameters \cite{Gales97}. The alignment of this model is used for neural network model training as the reference label. The multistream CNN AM described in Section 2.1 is trained on the total 960hr training set with the LF-MMI loss, decaying learning rates from $10^{-3}$ to $10^{-5}$ over the span of 6 epochs. The mini-batch size is 64.  

To prepare a lexicon, we select the most frequently used 200K words from the 800M word text and add out-of-vocabulary words to the original lexicon provided by the LibriSpeech corpus with the CMU phoneset, resulting in a 203K word list in total. We train
a G2P model using the Sequitur tool \cite{sequitur} to generate pronunciations for the out-of-vocabulary words. 

We use the PocoLM tooklit to train $n$-gram LMs by modifying the default recipe for the Switchboard corpus\footnote{https://github.com/danpovey/pocolm/blob/master/egs/swbd/run.sh.}. A 4-gram LM is trained on the 800M word text as well as the entire text transcripts for the 960hr training data containing around 10M words. This LM is pruned to a 3-gram, which is used for the 1st-pass decoding. The 4-gram LM is used for $n$-gram LM rescoring.

The TDNN-LSTM language model is trained on the aforementioned combined text, totaling 810M words, with the default Kaldi RNNLM recipe for LibriSpeech. We modify the dimension of embedding to 4,096 to increase the representational power of contexts in word sequences.

\subsection{Self-Attentive SRU Language Models}

\subsubsection{Dataset}
We construct our dataset for language modeling by combining the normalized corpus of 800M words with the text transcripts from the 960h training data, for a total of about 810M words. All of our self-attentivce SRU language models are trained at the utterance level (i.e., the model does not leverage any context past sentence boundaries), with a maximum sequence length of 275 tokens. We train a new 10K BPE vocabulary for our model. We limit the maximum sequence length only during training, not when computing dev set perplexity or when rescoring utterances from $N$-best hypotheses by the acoustic model with the TDNN-LSTM LM. We report perplexity numbers on dev-clean and dev-other, which include start and end of sentence tokens for each utterance.

\subsubsection{Model configuration}
All the self-attentive SRU LMs are trained using a hidden dimension of 2,048 and a projected dimension of 512 for the self-attention layer. We use a learning rate of $2\cdot 10^{-4}$ and no dropout. Optimization is done with the RAdam optimizer~\cite{liu2019radam} using a cosine annealing learning rate schedule. We train SRU models of 12 and 24 layers, slightly varying architectures. For the 12 layer model, we train across 8 Tesla V100 GPUs with a total batch size of 192 for 10 epochs. We use an embedding size of 2,048 and tie the input and output weights. We use single-headed attention in the self-attentive modules. We train the 24 layer model on 8 Quadro RTX 8000 GPUs with a total batch size of 512 for 12 epochs. We use an embedding size of 512, do not tie weights, and use 2 heads in the self-attentive modules.

\begin{table}[t]
\centering
\begin{tabular}{c|c|c|c|c}
\hline
\multirow{2}{*}{\textbf{Model}} & \multirow{2}{*}{\textbf{Layers}} & \multirow{2}{*}{\textbf{\# Params}} &\multicolumn{2}{c}{\textbf{Dev}} \\
\cline{4-5}
& & & clean & other \\
\hline
Transformer & 12 & 74M & 37.7 & 39.5 \\
\hline
SRU & 12 & 77M & 36.2 & 38.3 \\
\hline
SRU & 24 & 139M & 34.3 & 36.8 \\
\hline
\end{tabular}
\vspace{1em}
\caption{BPE-level perplexities on dev-clean and dev-other for 12 layer Transformer, 12 layer SRU and 24 layer SRU LMs. We include start and end of sentence tokens on each utterance.}
\label{table:lm_ppl}
\vspace{-5mm}
\end{table}

\subsubsection{Results}
Table \ref{table:lm_ppl} shows the perplexities obtained on the dev sets. With our 12 and 24-layer models, we achieve dev-clean perplexities of 36.2 and 34.3 respectively. In the first third of training we see the most improvement, with combined dev set perplexities at 40 for the 12-layer model and 37 for the 24-layer model.

For a comparison we also train a 12-layer Transformer model on the 10K BPE vocabulary. We use a model dimension of 768, feedforward dimension of 2,048, and 8 attention heads. These parameters were chosen as they result in a network with a comparable number of parameters to the 12-layer SRU model. All other parameters such as weight tying mirror the 12-layer SRU model. A cosine annealing learning rate schedule is used, with a linear warmup for the first 20,000 steps. By adding a self-attentive module between SRU layers, we are able to achieve better perplexity using a comparable number of parameters.

\subsubsection{Analysis}
\begin{figure}[t]
  \includegraphics[width=0.95\linewidth]{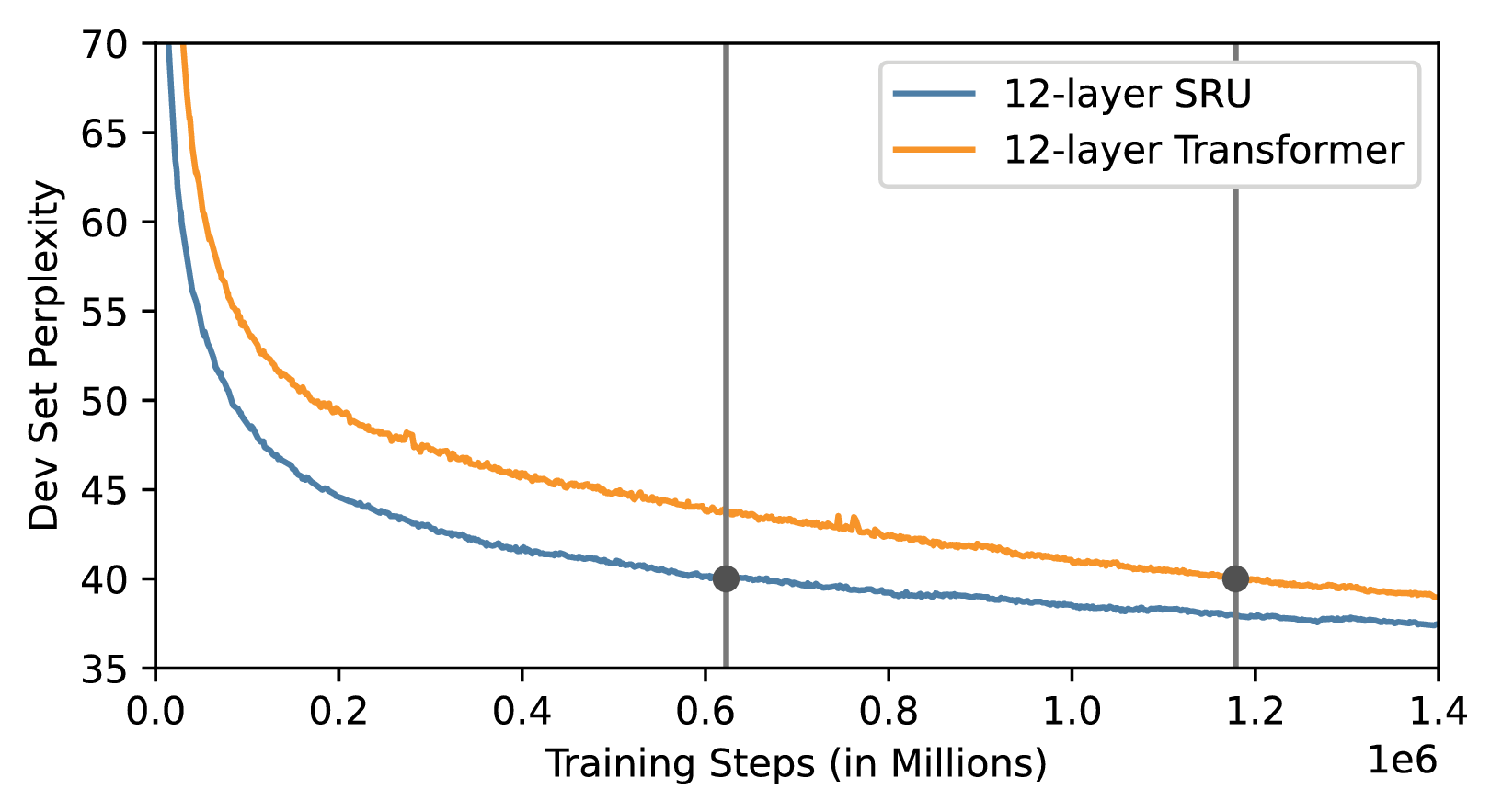}
  \caption{Dev perplexity curves of 12-layer SRU and Transformer models. Vertical lines signify when each model reaches a perplexity of 40. Here perplexity is reported on the combination of dev-clean and dev-other.}
  \label{fig:lm_ppl}
\end{figure}

We show that our proposed self-attentive SRU not only improves performance over the Transformer architecture but also converges faster. As shown in Figure \ref{fig:lm_ppl}, the 12-layer Transformer model reaches a perplexity of 40 in under 1.2M steps, while it only takes 622K training steps for the SRU model. This results in a 2 times training speedup. In practice this reduced training time by almost 2 days, allowing for faster iteration and greater exploration.

Additionally, we show that the perplexity improvement achieved by the 12-layer SRU model transfers directly to a WER improvement when used for candidate rescoring. In Table \ref{table:lm_wer} we compare WERs when using both the SRU model and Transformer for the final stage of $N$-best rescoring, fixing $N$ at 100. In all dev and test sets the 12-layer SRU achieves a lower WER than the Transformer model.

\begin{table}[t]
\centering
\begin{tabular}{c|c|c|c|c}
\hline
\multirow{2}{*}{\textbf{Rescoring Model}} & \multicolumn{2}{c|}{\textbf{Dev}} & \multicolumn{2}{c}{\small \textbf{Test}}\\
\cline{2-5} 
& clean & other & clean & other \\
\hline
12-layer Transformer & 1.62 & 4.41 & 1.96 & 4.70 \\
\hline
12-layer SRU & 1.59 & 4.38 & 1.93 & 4.62 \\
\hline
\end{tabular}
\vspace{1em}
\caption{WER (in \%) comparison of 12-layer Transformer and 12-layer SRU for $N$-best rescoring.}
\label{table:lm_wer}
\vspace{-5mm}
\end{table}

\subsection{Results and Analysis}
Table \ref{table:wer_setup} shows the WER comparison of different experimental setups for the multistream CNN AM and staged rescoring with various LMs. The setup of TDNN-F + 4-gram is a baseline with the TDNN-F acoustic model of the Kaldi Librispeech recipe rescored with our custom 4-gram LM. As compared to this basline, multistream CNN achieves a relative WER improvement of 14\% on test-other, demonstrating its robustness. The lattice rescoring with the TDNN-LSTM language model further reduces the WER by $14\%$ relative, showing the better modeling capability of a neural language model.

Regarding self-attentive SRU LMs, we first construct $N$-best rescoring using the 24-layer BPE SRU model. The language model likelihood is re-estimated by linearly interpolating the TDNN-LSTM and SRU LM. BPE-based LMs can help mitigate out-of-vocabulary issues from word-based models. Also, interpolating LMs with different levels of capacity has been proven to be beneficial to WER reduction in practice. These benefits are presented by the relative WER improvement of 23\% on the test sets against the TDNN-LSTM rescoring approach only. When we interpolate the BPE SRU model with the word-level SRU LM, we obtain a slight improvement around $2\%$ relative. Finally, we re-rank the interpolated SRU results by minimizing the expected WER, resulting in further reduction of WER by approximately $1\%$, also relative.

Table \ref{table:wer_system} compares the WERs between our proposed system and other benchmark systems in the literature. Other than the test-other set, we outperform any other system performances in the group by noticeable margins. In comparison with our previous results \cite{han19-2}, thanks to multistream CNN for acoustic modeling and multiple stages of LM rescoring with powerful self-attentive SRU language models, we improve the relative WERs on test-clean and test-other of 20\% and 23\%, respectively.

\begin{table}[t]
\centering
\begin{tabular}{l|c|c|c|c}
\hline
\hspace{8mm} \multirow{2}{*}{\textbf{Setup}} & \multicolumn{2}{c|}{\textbf{Dev}} & \multicolumn{2}{c}{\small \textbf{Test}}\\
\cline{2-5} 
& clean & other & clean & other \\
\hline
TDNN-F + 4-gram & 2.75 & 8.16 & 2.93 & 8.17 \\ 
\hline
Multistream CNN & \multirow{2}{*}{2.62} & \multirow{2}{*}{6.78} & \multirow{2}{*}{2.80} & \multirow{2}{*}{7.06} \\ 
+4-gram & & & &  \\
\hline
+TDNN-LSTM LM & 2.14 & 5.82 & 2.34 & 6.04 \\
\hline
+24-layer SRU & 1.56 & 4.28 & 1.83 & 4.57\\
\hline
+Interpolated SRU & 1.56 & 4.25 & 1.79 & 4.49 \\
\hline
+Expected Word & \multirow{2}{*}{1.55} & \multirow{2}{*}{4.22} & \multirow{2}{*}{1.75} & \multirow{2}{*}{4.46} \\ 
\hspace{1mm} Error Minimization & & & &  \\
\hline
\end{tabular}
\vspace{1em}
\caption{WER (in \%) comparison among different setups.}
\label{table:wer_setup}
\end{table}

\begin{table}[t]
\centering
\begin{tabular}{c|c|c|c|c}
\hline
\multirow{2}{*}{\textbf{Systems}} & \multicolumn{2}{c|}{\textbf{Dev}} & \multicolumn{2}{c}{\small \textbf{Test}}\\
\cline{2-5} 
& clean & other & clean & other \\
\hline
Park, et al. \cite{specaugment} & - & - & 2.5 & 5.8 \\
\hline
Synnaeve, et al. \cite{synnaeve19} & \multirow{2}{*}{2.10} & \multirow{2}{*}{4.79} & \multirow{2}{*}{2.33} & \multirow{2}{*}{5.17} \\ 
w/o semi-supervision & & & &  \\
\hline
Luscher, et al. \cite{Luscher2019} & 1.9 & 4.5 & 2.3 & 5.0 \\
\hline
Wang, et al. \cite{wang20} & - & - & 2.26 & 4.85 \\
\hline
Han, et al. \cite{han19-2} & 1.84 & 5.75 & 2.20 & 5.82 \\ 
\hline
Zhang, et al. \cite{zhang2020transformer} & - & - & 2.0 & 4.6 \\
\hline
Han, et al. \cite{han2020contextnet} & - & - & 1.9 & \textbf{4.1} \\
\hline
ASAPP-ASR & \textbf{1.55} & \textbf{4.22} & \textbf{1.75} & 4.46 \\
\hline
\end{tabular}
\vspace{1em}
\caption{WER (in \%) comparison among different systems.}
\label{table:wer_system}
\vspace{-2mm}
\end{table}

\section{Conclusions}
In this work, we proposed a hybrid ASR system that combines a novel acoustic model architecture, \emph{multistream CNN}, and an efficient language model, \emph{self-attentive SRU}. Through the multiple stages of LM rescoring and the expected word error minimization for $N$-best hypotheses re-ranking, we achieved a new state-of-the-art result on test-clean and competitive performance on test-other in the popular speech benchmark of Librispeech. Multi-resolution processing in a multistream architecture by multistream CNN manifested its robustness on test-other, and self-attentive variant to SRU demonstrated its superiority of modeling power over Transformer.

We will continue on improving the robustness of our acoustic model with efficient usage of a deep CNN architecture and more optimization of data augmentation methods in training. With the promising results presented by the self-attentive SRU in language modeling, we also plan to leverage similar modeling capacity from SRUs in acoustic modeling in the framework of end-to-end ASR.


\bibliographystyle{IEEEtran}

\bibliography{ref}

\begin{thebibliography}{10}
\providecommand{\url}[1]{#1}
\csname url@samestyle\endcsname
\providecommand{\newblock}{\relax}
\providecommand{\bibinfo}[2]{#2}
\providecommand{\BIBentrySTDinterwordspacing}{\spaceskip=0pt\relax}
\providecommand{\BIBentryALTinterwordstretchfactor}{4}
\providecommand{\BIBentryALTinterwordspacing}{\spaceskip=\fontdimen2\font plus
\BIBentryALTinterwordstretchfactor\fontdimen3\font minus
  \fontdimen4\font\relax}
\providecommand{\BIBforeignlanguage}[2]{{%
\expandafter\ifx\csname l@#1\endcsname\relax
\typeout{** WARNING: IEEEtran.bst: No hyphenation pattern has been}%
\typeout{** loaded for the language `#1'. Using the pattern for}%
\typeout{** the default language instead.}%
\else
\language=\csname l@#1\endcsname
\fi
#2}}
\providecommand{\BIBdecl}{\relax}
\BIBdecl

\bibitem{Hinton12}
G.~Hinton, L.~Deng, D.~Yu, G.~Dahl, A.~Mohamed, N.~Jaitly, A.~Senior,
  V.~Vanhoucke, P.~Nguyen, T.~Sainath, and B.~Kingsbury, ``Deep neural networks
  for acoustic modeling in speech recognition,'' \emph{IEEE Signal Process.
  Mag.}, vol.~29, no.~6, pp. 82--97, 2012.

\bibitem{Hochreiter97}
S.~Hochreiter and J.~Schmidhuber, ``Long short-term memory,'' \emph{Neural
  Comp.}, vol.~9, no.~8, pp. 1735--1780, 1997.

\bibitem{vaswani}
A.~Vaswani, N.~Shazeer, N.~Parmar, J.~Uszkoreit, L.~Jones, A.~N. Gomez,
  L.~Kaiser, and I.~Polosukhin, ``Attention is all you need,'' in
  \emph{NeurIPs}, 2017, pp. 5998--6008.

\bibitem{Xiong16-2}
W.~Xiong, J.~Droppo, X.~Huang, F.~Seide, M.~Seltzer, A.~Stolcke, D.~Yu, and
  G.~Zweig, ``Achieving human parity in conversational speech recognition,''
  2016, [Online]. Available: https://arxiv.org/abs/1610.05256.

\bibitem{William00}
J.~F. William, W.~M. Fisher, A.~F. Martin, M.~A. Przybocki, and D.~S. Pallett,
  ``{NIST} evaluation of conversational speech recognition over the telephone:
  {E}nglish and {M}andarin performance results,'' in \emph{NIST}, 2000.

\bibitem{simoyan14}
K.~Simonyan and A.~Zisserman, ``Very deep convolutional networks for
  large-scale image recognition,'' 2014, [Online]. Available:
  https://arxiv.org/abs/1409.1556.

\bibitem{He16}
K.~He, X.~Zhang, S.~Ren, and J.~Sun, ``Deep residual learning for image
  recognition,'' in \emph{CVPR}, 2016, pp. 770--778.

\bibitem{yu16}
D.~Yu, W.~Xiong, J.~Droppo, A.~Stolcke, G.~Ye, J.~Li, and G.~Zweig, ``Deep
  convolutional neural networks with layer-wise context expansion and
  attention,'' in \emph{Interspeech}, 2016, pp. 17--21.

\bibitem{Povey16}
D.~Povey, V.~Peddinti, D.~Galvez, P.~Ghahrmani, V.~Manohar, X.~Na, Y.~Wang, and
  S.~Khudanpur, ``Purely sequence-trained neural networks for {ASR} based on
  lattice-free {MMI},'' in \emph{Interspeech}, 2016, pp. 2751--2755.

\bibitem{Saon17}
G.~Saon, G.~Kurata, T.~Sercu, K.~Audhkhasi, S.~Thomas, D.~Dimitriadis, X.~Cui,
  B.~Ramabhadran, M.~Picheny, L.~Lim, B.~Roomi, and P.~Hall, ``English
  conversational telephone speech recognition by humans and machines,'' in
  \emph{Interspeech}, 2017, pp. 132--136.

\bibitem{Han17}
K.~J. Han, S.~Hahm, B.~Kim, J.~Kim, and I.~Lane, ``Deep learning-based
  telephony speech recognition in the wild,'' in \emph{Interspeech}, 2017, pp.
  1323--1327.

\bibitem{Xiong17}
W.~Xiong, L.~Wu, F.~Alleva, J.~Droppo, X.~Huang, and A.~Stolcke, ``The
  {M}icrosoft 2017 conversational speech recognition system,'' in
  \emph{MSR-TR-2017-39}, 2017.

\bibitem{Kurata17}
G.~Kurata, B.~Ramabhadran, G.~Saon, and A.~Sethy, ``Language modeling with
  highway {LSTM},'' in \emph{ASRU}, 2017.

\bibitem{han18}
K.~J. Han, A.~Chandrashekaran, J.~Kim, and I.~Lane, ``Densely connected
  networks for conversational speech recognition,'' in \emph{Interspeech},
  2018, pp. 796--800.

\bibitem{capio}
K.~J. Han, A.~Chandrashekaran, J.~Kim, and I.~R. Lane, ``Capio 2017
  conversational speech recognition system,'' 2018, [Online]. Available:
  http://arxiv.org/abs/1801.00059.

\bibitem{panayotov2015librispeech}
V.~Panayotov, G.~Chen, D.~Povey, and S.~Khudanpur, ``Libr{S}speech: {A}n {ASR}
  corpus based on public domain audio books,'' in \emph{ICASSP}, 2015, pp.
  5206--5210.

\bibitem{chan}
W.~Chan, N.~Jaitly, Q.~V. Le, and O.~Vinyals, ``Listen, {A}ttend and {S}pell:
  {A} neural network for large vocabulary conversational speech recognition,''
  in \emph{ICASPP}, 2016, pp. 4960--4964.

\bibitem{specaugment}
D.~S. Park, W.~Chan, Y.~Zhang, C.-C. Chiua, B.~Zoph, E.~D. Cubuk, and Q.~V. Le,
  ``Spec{A}ugment: {A} simple data augmentation method for automatic speech
  recognition,'' in \emph{Interspeech}, 2019, pp. 2613--2617.

\bibitem{Luscher2019}
C.~Lüscher, E.~Beck, K.~Irie, M.~Kitza, W.~Michel, A.~Zeyer, R.~Schlüter, and
  H.~Ney, ``{RWTH ASR} systems for {L}ibri{S}peech: {H}ybrid vs attention,'' in
  \emph{Interspeech}, 2019, pp. 231--235.

\bibitem{han19-2}
K.~J. Han, R.~Prieto, and T.~Ma, ``State-of-the-art speech recognition using
  multi-stream self-attention with dilated 1{D} convolution,'' in \emph{ASRU},
  2019, pp. 54--61.

\bibitem{synnaeve19}
G.~Synnaeve, Q.~Xu, J.~Kahn, T.~Likhomanenko, E.~Grave, V.~Pratap, A.~Sriram,
  V.~Liptchinsky, and R.~Collobert, ``End-to-end {ASR}: {F}rom supervised to
  semi-supervised learning with modern architectures,'' 2019, [Onlne].
  Available: https://arxiv.org/abs/1911.08460.

\bibitem{wang20}
Y.~Wang, A.~Mohamed, D.~Le, C.~Liu, A.~Xiao, J.~Mahadeokar, H.~Huang,
  A.~Tjandra, X.~Zhang, F.~Zhang, C.~Fuegen, G.~Zweig, and M.~L. Seltzer,
  ``Transformer-based acoustic modeling for hybrid speech recognition,'' in
  \emph{ICASPP}, 2020.

\bibitem{povey18tdnnf}
D.~Povey, G.~Cheng, Y.~Wang, K.~Li, H.~Xu, M.~Yarmohamadi, and S.~Khudanpur,
  ``Semi-orthogonal low-rank matrix factorization for deep neural networks,''
  in \emph{Interspeech}, 2018, pp. 3743--3747.

\bibitem{dauphin17}
Y.~N. Dauphin, A.~Fan, M.~Auli, and D.~Grangier, ``Language modeling with gated
  convolutional networks,'' in \emph{ICML}, 2017, pp. 931--941.

\bibitem{librilight}
J.~Kahn, M.~Rivière, W.~Zheng, E.~Kharitonov, Q.~Xu, P.-E. Mazaré,
  J.~Karadayi, V.~Liptchinsky, R.~Collobert, C.~Fuegen, T.~Likhomanenko,
  G.~Synnaeve, A.~Joulin, A.~Mohamed, and E.~Dupoux, ``Libri-{L}ight: {A}
  benchmark for {ASR} with limited or no uupervision,'' 2019, [Online].
  Available: http://arxiv.org/abs/1912.07875.

\bibitem{librivox14}
J.~Kearns, ``Libri{V}ox: {F}ree public domain audiobooks,'' \emph{Reference
  Reviews}, vol.~28, no.~1, pp. 7--8, 2014.

\bibitem{Tjandra20}
A.~Tjandra, C.~Liu, F.~Zhang, X.~Zhang, Y.~Wang, G.~Synnaeve, S.~Nakamura, and
  G.~Zweig, ``Deja-vu: Double feature presentation and iterated loss in deep
  {T}ransformer networks,'' in \emph{ICASSP}, 2020.

\bibitem{han2020contextnet}
W.~Han, Z.~Zhang, Y.~Zhang, J.~Yu, C.-C. Chiu, J.~Qin, A.~Gulati, R.~Pang, and
  Y.~Wu, ``Context{N}et: Improving convolutional neural networks for automatic
  speech recognition with global context,'' 2020, [Online]. Available:
  http://arxiv.org/abs/2005.03191.

\bibitem{han20}
K.~J. Han, J.~Pan, V.~K.~N. Tadala, T.~Ma, and D.~Povey, ``Multistream {CNN}
  for robust acoustic modeling,'' in \emph{Interspeech}, 2020, in review.

\bibitem{lei18}
T.~Lei, Y.~Zhang, S.~I. Wang, H.~Dai, and Y.~Artzi, ``Simple recurrent units
  for highly parallelizable recurrence,'' in \emph{EMNLP}, 2018.

\bibitem{Povey11}
D.~Povey, A.~Ghoshal, G.~Boulianne, L.~Burget, O.~Glembek, N.~Goel,
  M.~Hannemann, P.~Motlicek, Y.~Qian, P.~Schwarz, J.~Silovsky, G.~Stemmer, and
  K.~Vesely, ``The {K}aldi speech recognition toolkit,'' in \emph{ASRU}, 2011.

\bibitem{Li2018}
K.~Li, H.~Xu, Y.~Wang, D.~Povey, and S.~Khudanpur, ``Recurrent neural network
  language model adaptation for conversational speech recognition,'' in
  \emph{Interspeech}, 2018, pp. 3373--3377.

\bibitem{park2018fully}
J.~Park, Y.~Boo, I.~Choi, S.~Shin, and W.~Sung, ``Fully neural network based
  speech recognition on mobile and embedded devices,'' in \emph{NeurIPS}, 2018,
  pp. 10\,620--10\,630.

\bibitem{shangguan2019optimizing}
Y.~Shangguan, J.~Li, Q.~Liang, R.~Alvarez, and I.~McGraw, ``Optimizing speech
  recognition for the edge,'' 2019, [Online]. Available:
  https://arxiv.org/abs/1909.12408.

\bibitem{Koriyama_2020}
T.~Koriyama and H.~Saruwatari, ``Utterance-level sequential modeling for deep
  {G}aussian process based speech synthesis using simple recurrent unit,''
  \emph{ICASSP}, 2020.

\bibitem{stolcke1997explicit}
A.~Stolcke, Y.~Konig, and M.~Weintraub, ``Explicit word error minimization in
  $n$-best list rescoring,'' in \emph{Eurospeech}, 1997.

\bibitem{Gales97}
M.~J.~F. Gales, ``Maximum likelihood linear transformations for {HMM}-based
  speech recognition,'' \emph{Comp. Speech and Lang.}, vol.~12, pp. 75--98,
  1997.

\bibitem{sequitur}
M.~Bisani and H.~Ney, ``Joint-sequence models for grapheme-to-phoneme
  conversion,'' \emph{Speech Comm.}, vol.~50, no.~5, pp. 434--451, 2008.

\bibitem{liu2019radam}
L.~Liu, H.~Jiang, P.~He, W.~Chen, X.~Liu, J.~Gao, and J.~Han, ``On the variance
  of the adaptive learning rate and beyond,'' in \emph{ICLR}, April 2020.

\bibitem{zhang2020transformer}
Q.~Zhang, H.~Lu, H.~Sak, A.~Tripathi, E.~McDermott, S.~Koo, and S.~Kumar,
  ``Transformer {T}ransducer: A streamable speech recognition model with
  transformer encoders and {RNN-T} loss,'' in \emph{ICASSP}, 2020, pp.
  7829--7833.

\end{thebibliography}


\end{document}